# Spatially resolved rotation of the broad-line region of a quasar at sub-parsec scale[†]

GRAVITY Collaboration[•]: E. Sturm[1*], J. Dexter[1*], O. Pfuhl[1], M. R. Stock[1], R. I. Davies[1], D. Lutz[1], Y. Clénet[2], A. Eckart[3,4], F. Eisenhauer[1], R. Genzel[1,5], D. Gratadour[2], S. F. Hönig[6], M. Kishimoto[7], S. Lacour[2], F. Millour[8], H. Netzer[9], G. Perrin[2], B. M. Peterson[10,11,12], P.O. Petrucci[13], D. Rouan[2], I. Waisberg[1], J. Woillez[14], A. Amorim[15], W. Brandner[16], N. M. Förster Schreiber[1], P. J. V. Garcia[17,18], S. Gillessen[1], T. Ott[1], T. Paumard[2], K. Perraut[13], S. Scheithauer[16], C. Straubmeier[3], L. J. Tacconi[1], F. Widmann[1]

[1]Max Planck Institute for Extraterrestrial Physics, Giessenbachstr.1, 85748 Garching, Germany
[2]LESIA, Observatoire de Paris, Université PSL, CNRS, Sorbonne Université, Univ. Paris Diderot, Sorbonne Paris Cité, 5 place Jules Janssen, 92195 Meudon, France
[3]1. Physikalisches Institut, Universität zu Köln, Zülpicher Str. 77,50937 Köln, Germany
[4]Max Planck Institute for Radio Astronomy, Auf dem Hügel 69, 53121 Bonn, Germany
[5]Departments of Physics and Astronomy, Le Conte Hall, University of California, Berkeley, CA 94720, USA
[6]Department of Physics and Astronomy, University of Southampton, Southampton SO17 1BJ, UK
[7]Department of Physics, Kyoto Sangyo University, Motoyama, Kamigamo, Kita-ku, Kyoto 603-8555, Japan
[8]Université Côte d'Azur, Observatoire de la Côte d'Azur, CNRS, Laboratoire Lagrange, Nice, France
[9]School of Physics and Astronomy, Tel Aviv University, Tel Aviv 69978, Israel
[10]Department of Astronomy, The Ohio State University, 140 W 18th Ave, Columbus, OH 43210, USA
[11]Center for Cosmology and AstroParticle Physics, The Ohio State University, 191 West Woodruff Ave., Columbus, OH 43210, USA
[12]Space Telescope Science Institute, 3700 San Martin Drive, Baltimore, MD 21218, USA
[13]Univ. Grenoble Alpes, CNRS, IPAG, 38000 Grenoble, France
[14]European Southern Observatory, Karl-Schwarzschild-Str. 2, 85748 Garching, Germany
[15]CENTRA and Universidade de Lisboa − Faculdade de Ciências, Campo Grande, 1749-016 Lisboa, Portugal
[16]Max Planck Institute for Astronomy, Königstuhl 17, 69117, Heidelberg, Germany
[17]CENTRA and Universidade do Porto − Faculdade de Engenharia, 4200-465 Porto, Portugal
[18]European Southern Observatory, Alonso de Córdova 3107, Vitacura, Región Metropolitana, Chile
*These authors contributed equally to this work

**The broadening of atomic emission lines by high-velocity motion of gas near accreting supermassive black holes is an observational hallmark of quasars[1]. Observations of broad emission lines could potentially constrain the mechanism for transporting gas inwards through accretion disks or outwards through winds[2]. The size of this broad-line region has been estimated by measuring the light travel time delay between the variable nuclear continuum and the emission lines[3] – a method known as reverberation mapping. In some models the emission lines arise from a continuous outflow[4], whereas in others they are produced by orbiting gas clouds[5]. Directly imaging such regions has not hitherto been possible because of**

their small angular sizes (< 0.1 milli-arcseconds[3,6]). **Here we report a spatial offset (with a spatial resolution of ten micro-arcseconds or about 0.03 parsecs for a distance of 550 million parsecs) between the red and blue photo-centres of the broad Paschen-α line of the quasar 3C 273 perpendicular to the direction of its radio jet. This spatial offset corresponds to a gradient in the velocity of the gas and thus implies that the gas is orbiting the central supermassive black hole. The data are well fitted by a broad-line-region model of a thick disk of gravitationally bound material orbiting a black hole of $3 \times 10^8$ solar masses. We infer a disk radius of 150 light days; a radius of 100-400 light days was found previously using reverberation mapping[7-9]. The rotation axis of the disk aligns in inclination and position angle with the radio jet. Our results support the methods that are often used to estimate the masses of accreting supermassive black holes and to study their evolution over cosmic time.**

We observed the quasar 3C 273 at the Very Large Telescope Interferometer (VLTI) in Chile using the recently deployed instrument GRAVITY[10] on 8 nights between July 2017 and May 2018. The instrument coherently combines the light of the four 8 m telescopes to form interferometric amplitudes and phases on each of the 6 baselines (telescope pairs). Amplitudes measure the angular extent (size) of a structure, while phases provide its on-sky position. The continuum dust emission was partially resolved (diameter ~0.3 milli-arcseconds) and the broad line region was more compact. We extracted differential phase curves (interferometric phase as a function of wavelength, measured relative to the continuum) for each of the 6 baselines near the (cosmologically) redshifted Pa α line observed at λ ~ 2.17 micron and averaged them over time to increase the signal-to-noise. The differential phase (Δφ) measures the photo-center shift on sky (Δx) of the total (line+continuum) image along the projected baseline direction for an unresolved source:

$\Delta\phi(\lambda) = -2\pi \, f_{line} \, B/\lambda \, \Delta x(\lambda)$      (1),

where B is the sky-projected baseline length (telescope separation) and $f_{line}$ is the ratio of the emission line to total flux at each wavelength channel (Methods). In this way precise phase measurements from spectro-astrometry[11] provide spatial information on scales much smaller than the interferometric beam. 3C 273 is the most attractive target for spectro-astrometry. It is the brightest nearby quasar, with a large region size measured from reverberation mapping[7,8,9], and its strong Pa α line is observable in the near-infrared K band where GRAVITY operates.

The strength of the phase signal depends on the kinematics. Turbulent motion produces zero differential phase, while a spatial velocity gradient results in wavelength-dependent phase shifts. We detected such a velocity gradient signature on 3 of our 6 baselines (UT4-3, UT4-2, UT4-1, see Methods) that were not aligned with the 3C 273 jet direction. Averaging the data of those 3 baselines, we find phase peaks of 0.25 +/- 0.06 deg over multiple spectral channels (fig. 1a). We reject the null hypothesis of zero differential phase at > 5σ significance (Methods). This is a factor of > 10 higher precision in differential phase than previously achieved in interferometry of Active Galactic Nuclei (AGN)[12].

Using the differential phase data $\Delta\phi$ from all baselines, we measure the 2D photo-center position $\Delta x$ (model-independent image centroid) at each wavelength channel contributing significant line flux where equation 1 can be inverted ($f_{line} > 0.35$, 7 channels). We clearly detect a velocity gradient (fig. 1b, spatial separation between red and blue sides of the line), which is nearly perpendicular to the large-scale position angle of the 3C 273 radio jet[13] (222 deg). Fitting a model with symmetric photo-centers at the red and blue sides of the line to all of the data, the measured offset is $\Delta x = \pm (-9.5, 6.8) \pm (1.6, 1.1)$ micro-arcseconds in RA and Dec to the blue and red sides, corresponding to ~0.03 pc (~6,000 AU) in radius. This precision of ~1 micro-arcsecond is ~500 AU at a redshift of z = 0.158 (550 mega-parsecs).

A velocity gradient perpendicular to the jet is strong evidence of ordered rotation, although the signal-to-noise ratio in each channel is too low to uniquely determine the rotation profile. Here we use *ordered* in the sense of coherent motion producing a velocity gradient. Orbital motions with random angular momenta, for example, would not produce the phase signature that we detect.

To constrain the size and structure of the line-emitting region, we next adopt a kinematic model of a collection of non-interacting point particles ("clouds") with equal (arbitrary) line flux distributed in an annular thick disk geometry and assuming circular orbits in the black hole gravitational potential[14,15] (Methods). The 7 model parameters used are the mean and inner radii of the broad Pa $\alpha$ line emitting region (BLR) on sky ($R_{BLR}$, $R_{min}$), its radial shape parameter ($\beta$, ranging from concentrated around the mean to exponentially distributed), the disk opening angle ($\theta_o$), the observer's inclination and position angles (i, PA), and the central black hole mass ($M_{BH}$, see diagram fig. 1c). We find a good joint fit (reduced $\chi^2 \sim 1.3$) to the Pa $\alpha$ line profile and the interferometric phases on all baselines and 40 wavelength channels and constrain the model parameters using Markov Chain Monte Carlo sampling[16] (Methods). The centroids in each channel found by the model (dashed line in fig. 1b) are in good agreement with those observed. We infer a position angle of $210^{+6}_{-5}$ deg (with a 180 deg ambiguity), and an inclination of $12 \pm 2$ deg (all intervals at 90% confidence) for the rotation axis of the BLR. Measurements of the radio jet inclination angle from superluminal motion range from 7-15 deg[17,18,19]. The close 2D alignment of the rotation axis and the radio jet confirms that the kinematics are dominated by ordered rotation. The half opening angle of the gas distribution is $45^{+9}_{-6}$ deg.

The mean Pa $\alpha$ emitting region radius is found to be $R_{BLR} = 46 \pm 10$ μas ($0.12 \pm 0.03$ pc), with an inner edge at $R_{min} = 11 \pm 3$ μas ($0.03 \pm 0.01$ pc), and a roughly exponential radial emission profile with shape parameter $\beta = 1.4 \pm 0.2$ (Methods). The measured mean radius corresponds to $145 \pm 35$ light days, about a factor of 2 smaller than previous reverberation mapping estimates (260-380 light days) using H$\beta$ and H$\gamma$ emission lines[7,8] but consistent with a lower limit > 100 light days found from a later re-analysis[9]. The discrepancy is likely due to the difficulty in measuring long lags in the brightest quasars, and could partially be due to intrinsic source variability (see Methods).

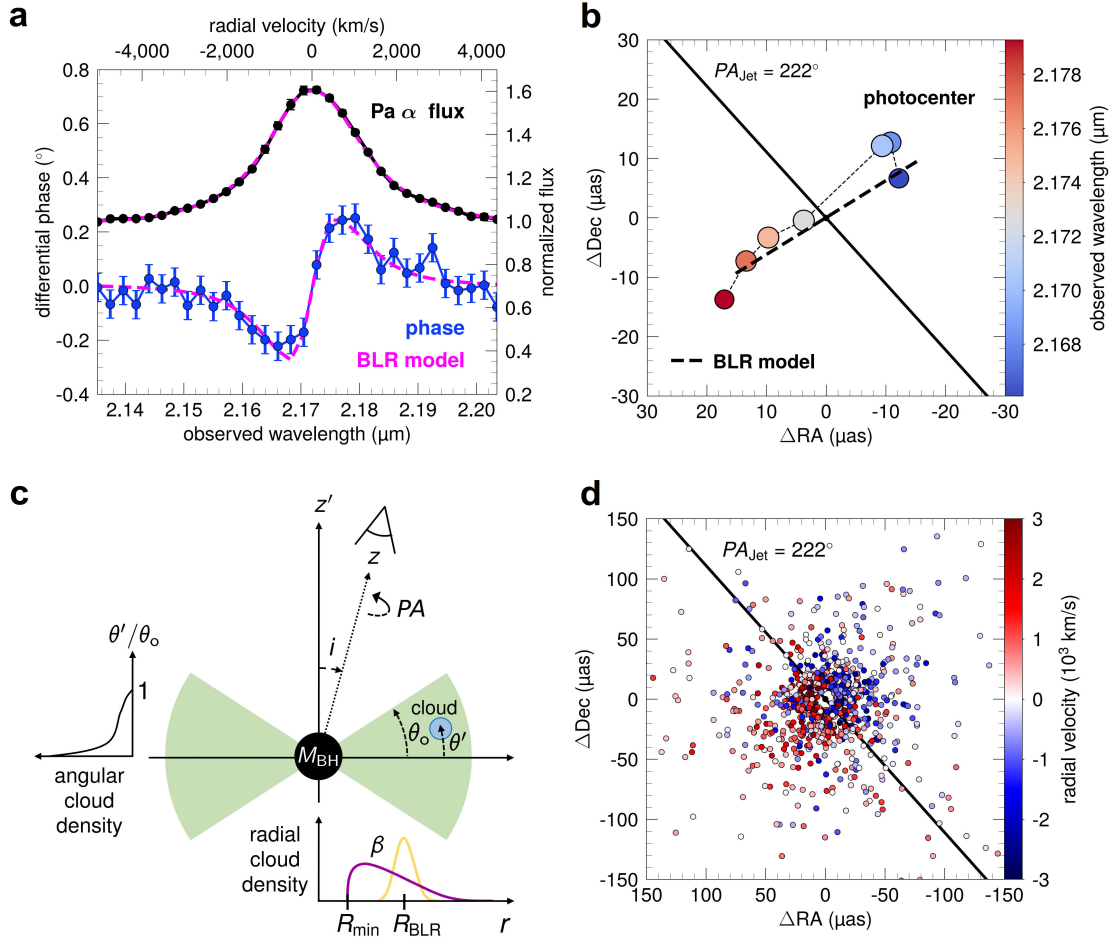

**Figure 1: Main observational and modelling results**. **(a)** Observed GRAVITY 3C 273 Pa α line profile (black points) and differential phases averaged over three baselines (blue points) showing the S-shape typical for a velocity gradient. The error bars represent 1 sigma uncertainties. A thick disk model (pink lines, see also panels c and d) of the broad line emitting region provides an excellent joint fit to the data. **(b)** Observed centroid positions at several wavelength channels (symbol size proportional to signal-to-noise) show a clear spatial separation between red- and blue-shifted emission: a velocity gradient at a position angle nearly perpendicular to that of the radio jet (solid black line). The centroid track of the model is shown as the dashed line. **(c)** Schematic representation of the model parameters. **(d)** Velocity map of the best fitting model, a thick disk geometry viewed nearly face-on. Disorder in the velocity map reduces the observed photo-center shifts (b) compared to the angular size (d) of the broad Pa α emitting region.

Our inferred BLR radius is also a factor ~3 smaller than the continuum dust radius found from past interferometry measurements[20] and our own data. This has been found for many AGN[21]. The size is also much smaller than that tentatively found from previous spectro-interferometry of 3C 273 using the VLTI instrument AMBER[12] ($R_{BLR}$ ~ 200 μas). The object was too faint for fringe tracking in those observations. In addition, one of the three baselines used by AMBER was in the direction of the jet, i.e. perpendicular to the disk. Our GRAVITY data at ~40x higher precision rule out such a large size (fig. E4).

The inferred inner edge of the Pa α emission region is a result of the cutoff in the line profile at ± 4,000 km/s, likely corresponding to the location where Balmer and Paschen emission becomes weak compared with that of higher ionization lines[22]. The best fitting structure is similar to that found from velocity-resolved reverberation mapping of nearby Seyfert 1 galaxies[23,24], suggesting that BLR properties may not vary strongly with AGN luminosity or Eddington ratio.

We infer the black hole mass of 3C 273 directly from the model as $M_{BH}$ = (2.6 ± 1.1)x$10^8$ solar masses. In reverberation mapping experiments, $M_{BH}$ is obtained by combining Balmer line time delay measurements with the gas velocity obtained from the line profile. This requires the use of a velocity-inclination factor $f = GM_{BH}/v^2R_{BLR}$, where v is the gas velocity usually defined as either $\sigma_{line}$, the second moment of the line profile, or $FWHM_{line}$. Empirical mean values of $< f >$ are obtained by matching reverberation mapping mass estimates with those from the M-$\sigma$ relationship[25,26]. Typically $< f > = 4.3$ for v = $\sigma_{line}$ and $< f > = 1.1$ for v = $FWHM_{line}$. For the observed 3C 273 Pa α line, we measure $\sigma_{line}$ = 1,400 km/s (f = 4.7 ± 1.4) and $FWHM_{line}$ = 2,700 km/s (f = 1.3 ± 0.2), both in good agreement with the mean values used for Hβ in the literature[27].

Because they used a larger $R_{BLR}$ = 260-380 light days[7,8,27] for 3C273, past studies found a factor ~2 higher mass. The mass is correlated with inclination: higher mass requires lower inclination in order to match the observed line width. Our $R_{BLR}$ and $M_{BH}$ could also in principle be underestimated: a larger BLR with added disorder may also be able to explain the phase signatures we observe. Our highest velocity photo-center on the blue side (which has the lowest signal-to-noise and is therefore not shown in figure 1) is offset (1-2σ) to the South and lies along the jet (see Methods, figure E2). This could be a hint of a second kinematic component contributing line flux.

Our results support the fundamental assumptions of reverberation mapping. The broad line width is dominated by bound motion in the black hole gravitational potential, and our inferred values of f match what is commonly found from the mean population. The region size is a factor ~2 smaller than what is typically assumed, but within the large range of reported results for 3C 273. Quantitative size comparisons should be done using Seyfert galaxies with shorter lags where the radius is better determined from reverberation mapping. A comparison between velocity-resolved reverberation and (contemporaneous) interferometry in the same object would be particularly promising: reverberation mapping can unambiguously distinguish rotation from inflow and outflow, interferometry measures the orientation on sky, and both techniques independently probe size and structure. The thick disk structure that we infer in 3C 273 could either have a physical origin in the inflowing surface layer of a geometrically thick accretion disk[28] or in a rotating wind launched from the surface of a thin disk[29]. Distinguishing those scenarios will require detailed model predictions, since strong outflows can still show dominant signatures of rotation[30].

The combination of low inclination and large opening angles, small radius, and alignment of our baselines with the rotation axis reduces the observed phase amplitude (~0.3 deg) from the predicted value (> 1 deg). Nonetheless, we have shown that the remarkable sensitivity of GRAVITY allows measurements of BLR size and structure. Future observations could aim to measure a rotation curve and search for additional (e.g. outflow) components and/or structural variability. With current instrumentation spectro-interferometry is limited to the brightest AGN (K <~ 12, V <~ 15). With an ongoing observing programme, we plan to measure line and torus emission region properties in a small sample of quasars and Seyfert galaxies[10]. The angular size of each component scales with the observed optical flux[6], making interferometry particularly well suited to exploring the physical origin of the broad line region and measuring Black Hole masses in luminous quasars like 3C 273 which are more representative of the supermassive black holes found in large samples out to high redshift.

**Table 1 | Estimates of the kinematic BLR model parameters**

| Parameter | Value | Description |
|---|---|---|
| $R_{\mathrm{BLR}}$ (μas) | 46 ± 10 | Mean angular distance of the cloud from the black hole |
| $R_{\mathrm{min}}$ (μas) | 11 ± 3 | Minimum angular distance of the cloud from the black hole |
| $\beta$ | 1.4 ± 0.2 | Radial distribution shape parameter |
| $\theta_{\mathrm{o}}$ (°) | $45^{+9}_{-6}$ | Half-opening angle of the disk |
| $i$ (°) | 12 ± 2 | Inclination angle of the observer |
| PA (° E of N) | $210^{+6}_{-9}$ | Position angle on the sky |
| $M_{\mathrm{BH}}$ ($10^8 M_{\odot}$) | 2.6 ± 1.1 | Black-hole mass |

Best values are medians and uncertainties represent 90% confidence ranges

**Extended Data**     Figures E1,2,3,4 and Table E1

**Acknowledgements** We are grateful to ESO and the ESO/Paranal staff, and to the many scientific and technical staff members in our institutions who helped to make GRAVITY a reality. J.D. and M.R.S. were supported by a Sofja Kovalevskaja award from the Alexander von Humboldt foundation. S.F.H. acknowledges support from the EU Horizon 2020 ERC Starting Grant *DUST-IN-THE-WIND* (ERC-2015-StG-677117). M.K. acknowledges support from JSPS (16H05731). P.O.P. acknowledges financial support from the French PNHE. A.A and P.J.V.G. acknowledge support from FCT-Portugal with reference UID/FIS/00099/2013. J.D. thanks Anna Pancoast for useful discussions related to velocity-resolved reverberation mapping models and observations. E.S. thanks Leonard Burtscher for discussions in the preparatory phase of the project.

**Author Contributions** E.S., J.D., O.P., R.I.D., D.L., Y.C., F.E., R.G., D.G., S.F.H., M.K., F.M., H.N., G.P., B.M.P., P.O.P., D.R. and I.W. designed the GRAVITY AGN open time program; E.S., J.D., O.P., R.I.D., F.E., G.P., and I.W. planned and performed the observations; J.D., M.R.S. and O.P. processed, analyzed and modelled the data; J.D., E.S., M.R.S., O.P. and H.N. drafted text, figures and methods. All authors helped with the interpretation of the data and the manuscript.

**Author Information** Reprints and permissions information is available at www.nature.com/reprints. The authors declare no competing financial interests. Correspondence and requests for materials should be sent to E.S. (sturm@mpe.mpg.de) and/or J.D. (jdexter@mpe.mpg.de).

## Methods

### Observations

The observations were taken at the VLTI in Chile using the second-generation instrument GRAVITY[10] and the four 8 m Unit Telescopes. We chose the medium resolution mode of GRAVITY with 90 independent spectral elements (R~500) dispersed across 210 pixels. All data were obtained in single-field on-axis, combined polarization mode.

Each observation followed the same sequence. Each UT locked its adaptive optics (MACAO) module on 3C 273. Once the AO loops were closed, the telescope beams were coarsely aligned on the VLTI lab camera IRIS. In a second step the GRAVITY internal beam tracking aligned the fringe-tracking and science fibers on the target. The science observations were multiple blocks of 10 times 30s exposures (NDIT=10 and DIT=30s). Fringe tracking on such faint sources like 3C 273 (K~10, V~13) is difficult, so the strategy was to integrate deeply on source taking only occasional sky exposures, often at the end of the integration. In July 2017, Jan 2018, and May 2018 interferometric calibrators were also observed (although not required for differential phase measurements). Table E1 lists the observing nights, the integration time, the seeing, and the coherence time.

### Data reduction

We used the standard GRAVITY pipeline to process the data[31,10]. Each exposure consists of 10 science frames, which are averaged after processing. Each frame is corrected for the background bias by subtracting the closest sky exposure. The detector noise bias is calculated from dark exposures with the same settings as the science exposures. Flat-field and wavelength calibration are done on the internal calibration source. The effective wavelength for each spectral element is calibrated by modulating the fiber lengths of the fringe-tracking (FT) and science channel (SC) and using the internal laser metrology as a wavelength reference.

The science data are reduced using a pixel-to-visibility matrix[32] (P2VM). The P2VM represents the matrix-encoded instrument transfer function, which includes the relative throughput, coherence, phase-shift and cross-talk for each pixel. Applying the matrix to the detector frames yields the instrument-calibrated complex visibilities. In a second step the SC complex visibilities are phase-referenced to the FT complex visibilities using the laser metrology. The effective optical-path-difference for each spectral channel is calculated based on the delay measured by the laser metrology and the differential dispersion between the laser and the channel wavelength. This yields phase-referenced complex visibilities. The GRAVITY pipeline removes a mean and slope from the raw visibility phase calculated using all wavelength channels to create a differential phase on each baseline.

We used an alternative method (developed for VLTI/AMBER[33]) and computed the differential phase of each channel as in ref. 33 equation (7), which ignores the work channel in calculating the mean and slope. This method produces consistent results but improves our phase errors typically by ~10-20%. In order to account for the observatory transfer function (coherence loss due to vibrations, uncorrected atmosphere, birefringence, etc.) it is common practice to observe a calibrator star close to the science target. We have done this for 3 epochs in order to obtain continuum visibilities.

**Data processing**

For each exposure, we further removed broad instrumental shapes in the differential phases. This was done by convolving the differential phase vs. wavelength from each baseline with a Gaussian of FWHM 24 pixels, and then dividing the actual data by this smoothed version. The width was chosen to flatten the differential phases without affecting their behavior near the Pa α line. After flattening, no calibrator shows any significant differential phase signature at any part of the spectrum, down to limits < 0.1 deg. Therefore no further instrumental bias needs to be removed from the 3C 273 data. From each night, we have selected exposures where: i) the fringe tracking was working > 80% of the time, and ii) the rms of the differential phase curves was < 3 deg on all baselines. This selection removed 25% (26/104) of the 5-min exposures. Table 1 summarizes the data used in the analysis presented here.

We then averaged the exposures in time for each night for each of the 4 epochs to reduce the phase noise. For the first 3 epochs, the changes in (u,v) coordinates for each baseline are small during integrations of < 1 hr. In May 2018, the integrations are longer. We have checked for differences between the first and second half of each night, as well as the expected changes in phase signatures in the best fitting model (see below). In both cases, the changes are smaller than our errors and so we consider it safe to time-average per baseline at all epochs. The weights for each exposure and baseline were calculated from the rms of the phase noise over a broad spectral region centered on the Pa α line (2.06-2.28 micron). The averaged differential phase curves have a residual scatter of 0.2-0.3 deg in each epoch. Individual 5-min exposures typically reached 0.7-1.0 deg. In each epoch the precision reached is a factor ~10 higher than past AGN spectro-interferometry[12].

We extracted the spectral line profile in each exposure by summing the photometric flux of each of the 4 telescopes. We removed the shape of the instrument response by dividing the 3C 273 spectrum by that of a calibrator star. Spectral lines from the calibrator star were removed by first dividing its spectrum by templates from the NASA Infrared Telescope database[34]. For March/April 2018 where no calibrator was taken, we used that from the January 2018 observation instead. Then the red continuum slope was removed before averaging the result over all exposures within each epoch. The line shape and strength were stable between epochs to within 5%. That scatter could be either due to intrinsic variability or systematic error in the spectral extraction. Conservatively we assumed that the line profile and flux are constant and averaged the 4 epochs to get a single line profile. The errors are taken as the quadrature sum of the rms between epochs and the statistical errors on the measured photometric flux.

**Detection of a velocity gradient**

As described in the main text, we first used the differential phase data to fit for a best centroid position at each spectral channel across the line. The model has an (x,y) position with a phase given by $\Delta\phi_{ij} = -2\pi f_{line} (u_j x_i + v_j y_i)$, where the index i corresponds to the 7 spectral channels used (where the line intensity relative to the continuum $f_i > 0.35$ with $f_{line} = f_i/(1+f_i)$) and j corresponds to each of the 24 baselines (6x4 epochs). This is the form in the marginally resolved limit, where we can expand the exponential in the complex visibility and keep only the first order term[35,36]. In each channel we minimize the fit to the observed phases to find a best centroid position $x_i$. The results are shown in figure E2, where the red and blue wavelength channels clearly cluster on opposite sides of the line. The 1σ confidence intervals are shown as ellipses.

To estimate the significance of the detection, we consider a null hypothesis of zero phase everywhere and compare this to a model with a single centroid position (x,y) and (-x,-y) in RA and Dec on the blue and red sides of the line. We calculate the model phases as above for channels whose centroids appear to deviate significantly from zero (4 red and 4 blue channels) and assign them zero phase elsewhere. From least squares fitting, we find $(x, y) = (-9.5, 6.8) \pm (1.1, 1.6)$ µas. The $\chi^2$ for the null hypothesis and model are 1,417 and 1,308 with 960 data points. An F-test rejects the x=y=0 null hypothesis with a p-value of $10^{-17}$, corresponding to $> 8\sigma$. The spectral channels are more finely sampled than their resolution element, so that neighboring channels are correlated. To estimate the impact of this effect we repeat the above test using half of the channels and find a p-value for the null hypothesis of $< 10^{-8}$ or a detection significance of $> 5.5\sigma$.

The bluest channel in figure E2 moves to the South, along the jet. Its phase agrees with those of other blue channels for the 3 "off jet" baselines (figure 1a). However, it shows a -0.2 deg phase in the average of the other 3 baselines. The significance is only 1-2$\sigma$, and we do not interpret it further here. If detected in future observations, the alignment with the large-scale jet direction could be a signature of outflowing gas at high velocity. At certain baselines and spectral channels we see apparently systematic features that can extend over 2-3 channels, away from any spectral lines. None of these reach close to the signal-to-noise of the signature studied here. This is true for all GRAVITY AGN and calibrator data examined.

**BLR model description and fitting procedure**

We adopt a phenomenological model of the Pa $\alpha$ emitting region to interpret our data closely following past work[14,15]. The model comprises a large number of non-interacting test particles on circular orbits around the central black hole (BH) under the sole influence of gravity. The particles represent dense, low filling factor, line-emitting gas clouds. Their distances from the black hole are given by,

$$r = R_S + F\, R_{BLR} + g\,(1 - F)\, \beta^2\, R_{BLR}, \qquad (1)$$

where $R_S = 2GM_{BH}/c^2$ is the Schwarzschild radius, $R_{BLR}$ is the mean radius, $F = R_{min}/R_{BLR}$ is the fractional inner radius, $\beta$ is the shape parameter, and $g = p(x\,|\,1/\beta^2, 1)$ is drawn randomly from a Gamma distribution,

$$p(x\,|\,\alpha, \theta) = x^{\alpha-1} \exp(-x/\theta)\,/\,[\theta^\alpha\, \Gamma(\alpha)]. \qquad (2)$$

No line-emitting clouds are present inside of $R_{min}$ ($R_{min} >> R_S$ for the viable models found for 3C 273), and their distribution is allowed to vary from a Gaussian distribution concentrated around the mean ($0 < \beta < 1$) to an exponential ($\beta = 1$) or steeper profile ($1 < \beta < 2$) concentrated at the inner radius. The angular distribution is specified by a half-thickness $\theta_o$, and the clouds are placed at random positions along their orbits. The structure is viewed at an inclination angle i and rotated in the sky plane by a position angle PA measured in degrees E of N. In total the model as implemented has 7 free parameters: $R_{BLR}$, F, $\beta$, $M_{BH}$, $\theta_o$, i, PA.

Recent velocity-resolved reverberation mapping studies used additional parameters describing inflow and outflow, anisotropic emission, and asymmetry in the angular distribution[23,24]. Here we choose the minimal model required by our data. Rotation explains the velocity gradient perpendicular to the jet axis, an inner radius is justified by cutoffs in the spectrum at roughly +/- 4,000 km/s, and steep radial and thick angular distributions can produce the observed line shape. While we find a satisfactory fit (see below), including elliptical and/or radial motion would likely somewhat change our inferred parameters. In particular, additional model components that add disorder could potentially allow models with larger $R_{BLR}$ and $M_{BH}$ to fit the data.

Assuming that emitted Pa $\alpha$ photons free-stream through the BLR once emitted, we conveniently obtain spectro-interferometric observables as sums over clouds binned by the observed wavelength,

$$\frac{\lambda_{obs}}{\lambda_{emit}} = \left(1 + \frac{v_z}{c}\right)\left(1 - \frac{v^2}{c^2}\right)^{-1/2}\left(1 - \frac{R_S}{r}\right)^{-1/2}, \qquad (3)$$

for emitted wavelength $\lambda_{emit}$, total velocity v, and line-of-sight velocity $v_z$. We have included the relativistic and transverse Doppler shift and the gravitational redshift, since these effects could impact the emission line shape[37]. We note that equation 18 of Pancoast et al.[14] neglects the transverse Doppler shift, which is of the same order as the gravitational redshift for orbital motion. We also note that in equation (3) $v_z < 0$ corresponds to motion towards the observer (blueshift) in keeping with the standard radial velocity convention.

The spectral line shape is then found by summing the clouds in bins of observed wavelength. We account for the shape of the GRAVITY line spread function in medium resolution by binning at higher spectral resolution before convolving with a Gaussian of 4 nm FWHM. We then normalize the line strength so that it matches the observed strength at the peak and shift it to the observed central wavelength. The scaling and shift are both fixed in the analysis.

We model the differential phase of the BLR relative to a continuum that is assumed to be symmetric, as implied by < 1 deg observed closure phases in 3C 273. Since the region size (< 100 µas) is much less than the VLTI imaging resolution (~3 mas), each cloud contributes a phase of $\phi_k(\lambda, u, v) = -2\pi f(\lambda)/[1 + f(\lambda)](u\,x_k + v\,y_k)$ in radians, where $f(\lambda)$ is the line intensity relative to the continuum and (u,v) is the baseline separation. The Fourier Transform is linear, so we can find the total phase for each baseline by summing the individual phases in wavelength bins, using the same procedure as above to account for the instrument spectral PSF.

We fit the model with 7 parameters to all observed 3C 273 spectral line and phase curves simultaneously (40 wavelength channels for the time-averaged spectrum and 24 baselines, data shown in figure E1). The number of wavelength channels is chosen to fully cover the line profile as well as a small off region, since the inner radius we infer depends on the maximum observed radial velocity in the tails of the line. We use $2 \times 10^5$ clouds in the model, the minimum number so that the likelihood does not vary significantly between random instantiations. We use Bayesian statistics to measure confidence intervals on the model parameters. The priors are uniform in $\log\left(\frac{R_{BLR}}{\mu as}\right) \in (0,4)$, $F \in (0,1)$, $\beta \in (0,2)$, $\log\left(\frac{M_{BH}}{M_\odot}\right) \in (6,10)$, $\theta_0(\deg) \in (0,90)$, i (deg) $\in (0,90)$, PA (deg) $\in (0,360)$. The posterior is sampled using the emcee Markov Chain Monte Carlo code[16], with a joint likelihood including both the spectral and differential phase data across all spectral channels.

The sampling is run with 1,000 "walkers" independently exploring the parameter space, and converges within about 800 trials (a total of $8 \times 10^5$ samples of the likelihood). We have checked the convergence by comparing confidence intervals taken from well separated trials, and by starting the fit from several initial guesses far from the final distribution. We report 90% confidence intervals taken as the 5th and 95th percentiles of samples in the 1D distribution over each parameter. A corner plot of the full set of 2D and 1D distributions is shown in figure E3.

**Inferred BLR size compared with reverberation mapping**

Past reverberation mapping studies using Hβ and Hγ have found time lags of 260-380 light days[7,8] or a lower limit of > 100 light days[9]. Our measured value of 145 +/- 35 light days using Pa α is within this range, but a factor of 2 smaller than what is typically assumed. The wide range in the reported values is due to the difficulty of reverberation mapping measurements for luminous quasars which are equatorial on sky like 3C 273. The light curves must be combined across long seasonal observing gaps, leading to aliasing and other problems with irregular sampling. We believe that this is the main contributor to the discrepancy. Additionally, about 25% of this difference can be explained by the half a magnitude higher source luminosity during the past observations. Other contributing factors could be a real difference between Pa α and the Balmer line emission regions, but this seems unlikely since those lines share an upper atomic level. Our measured line width is slightly smaller than what has been measured for the other lines, implying a larger radius, but this could also be the result of complex unaccounted for kinematics.

As discussed in the main text, our measured size is consistent with the other known properties of 3C 273. It is a factor ~3 smaller than the hot dust continuum, similar to what is seen in other sources[21]. However, it is well below the prediction from the radius-luminosity relation based on Hβ (260-380 light days)[38]. Other recent reverberation mapping work on samples of more luminous sources find smaller source sizes and a flatter radius-luminosity relation[39,40], which would agree with our size estimate. Additional interferometry measurements, well suited to luminous quasars, can help to study this further.

**Limits on radial inflow and outflow**

With only detections of differential phases, spectro-interferometry cannot distinguish between rotation and ordered radial motion (inflow or outflow)[15]. Here we see that the velocity gradient is nearly perpendicular to the large-scale jet, and conclude that it indicates rotation. In a mixed model where both rotation and inflow/outflow are present, the position angle of the velocity gradient changes depending on their relative strength. As a simple estimate of the allowed fraction of inflow and outflow, we assume that the true rotation axis aligns with the 3C 273 jet. The observed offset between the two would then be due to outflow, which moves the net velocity gradient towards the jet direction. Fitting our previous BLR model with a fixed PA constrains the fraction of radial outflow at the local escape speed to be <~ 25%. This limit is qualitative, and for example changes depending on the true PA of the rotation axis and the implementation of the outflow model. In particular, disk wind models are often rotation dominated at low velocity[30] and would likely not be subject to this constraint.

**Size comparison with past 3C 273 near-infrared spectro-interferometry**

Reference 12 found tentative evidence for a very large BLR ($R_{BLR} \sim 200$ μas) using VLTI/AMBER. This result came from differential visibility amplitude data, where a claimed decrease in the wavelength-dependent change of the interferometric amplitude implies a larger image size of the line than of the continuum[35,36]. We show the differential amplitude data from our GRAVITY observations in figure E4, averaged over our two longest baselines (UT4-1, UT3-1) and all epochs. At >10x higher precision, we see no sign of the claimed decrease at the line as previously reported[12].

Instead we see a clear increase in the amplitude at the line, showing that the line is more compact than the hot dust continuum emission, $R_{BLR} \ll 150 \mu as$[20]. This compact size agrees with the small $R_{BLR}$ we independently infer from modeling the Pa $\alpha$ spectral line shape and the differential visibility phases as described above.

## Additional references

**Data Availability** The data discussed here are available on the archive of the European Southern Observatory (http://archive.eso.org/eso_archive_main.html).

# Extended data
## Table E1: GRAVITY data used for this work

| Date | Mode<br><br>Resolution/Polarisation | $t_{exp}$ on 3C 273<br><br>(min) | DIMM Seeing<br><br>(") | Coherence time<br><br>Tau$_0$ (ms) |
|---|---|---|---|---|
| 2017 Jul 07 | MED/COMBINED | 40 | 0.44 – 0.77 | 4.6 – 6.5 |
| 2017 Jul 08 | MED/COMBINED | 35 | 0.45 – 0.58 | 2.8 – 3.8 |
| 2018 Jan 08 | MED/COMBINED | 40 | 0.44 – 0.59 | 6.9 – 9.0 |
| 2018 Mar 29 | MED/COMBINED | 25 | 0.41 – 0.48 | 11.6 – 15.0 |
| 2018 Apr 01 | MED/COMBINED | 30 | 0.59 – 1.0 | 3.7 – 4.7 |
| 2018 Apr 02 | MED/COMBINED | 30 | 0.46 – 0.85 | 3.2 – 4.9 |
| 2018 May 29 | MED/COMBINED | 100 | 0.52 – 0.75 | 3.1 – 5.8 |
| 2018 May 30 | MED/COMBINED | 90 | 0.48 – 0.68 | 2.9 – 4.1 |

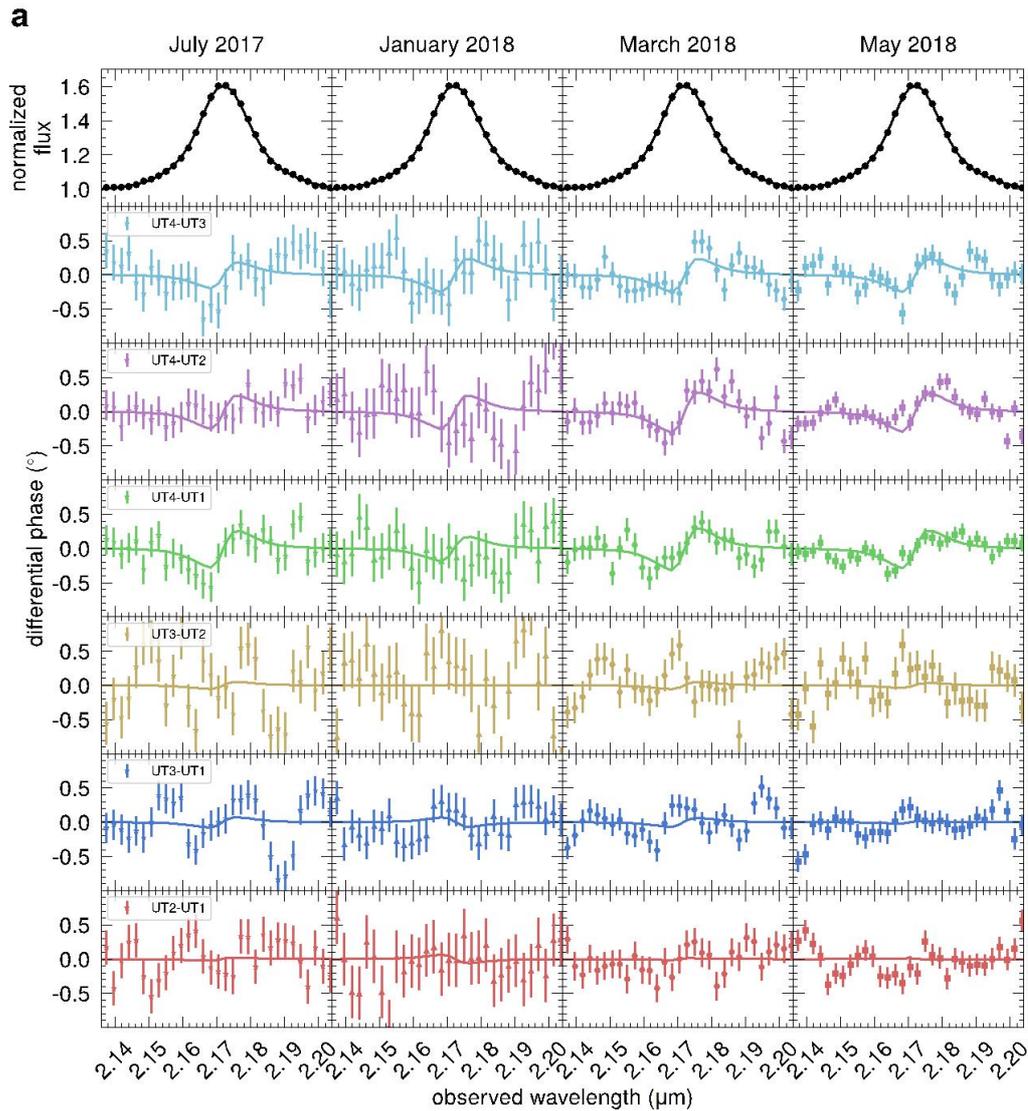

**a**

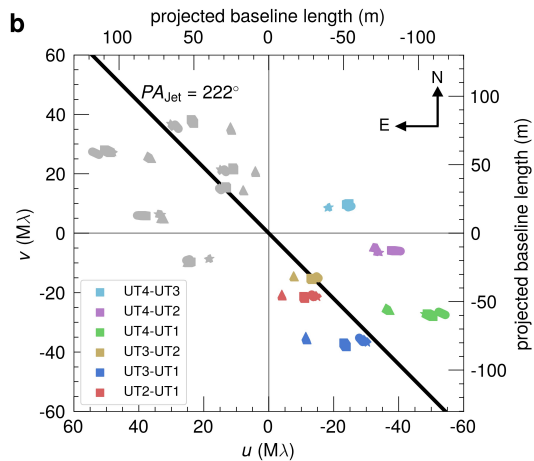

**b**

**Figure E1: Differential phases and uv-coverage. (a)** Differential phase curves (colored points with 1 sigma error bars) on the 6 VLTI baselines at 4 epochs (from left to right July 2017, Jan 2018, Mar 2018, May 2018) and time-averaged 3C 273 Pa α line profile (black points, identical in each panel). The best fitting BLR model to all of these data is shown as the solid lines. **(b)** uv-coverage for all of our data. Note the close alignment between all baselines, and particularly those without UT4, with the PA of the 3C 273 large-scale radio jet.

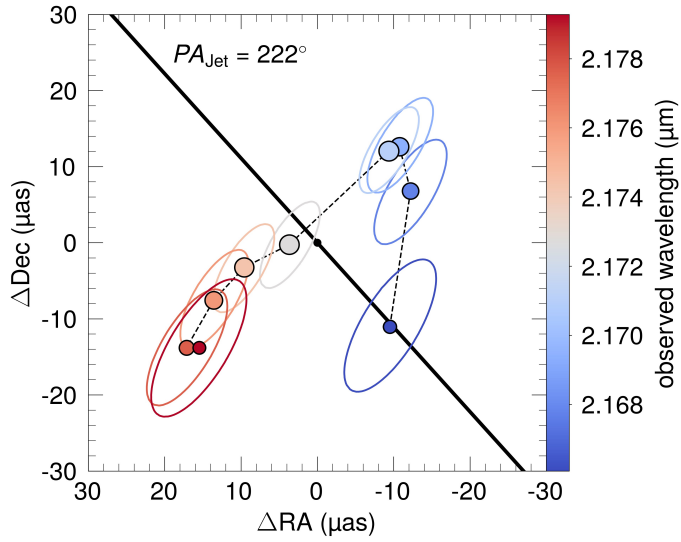

**Figure E2: Observed centroid positions at several wavelength channels.** Best fitting centroids to the differential phase data at each wavelength channel as in figure 1, but with contour ellipses containing 68% of the probability density. The extremal points to the blue (on the jet axis) and red are not shown in figure 1, because of their larger errors. Given the relatively low signal-to-noise in each channel, we cannot measure the radial velocity as a function of position (e.g. rotation curve).

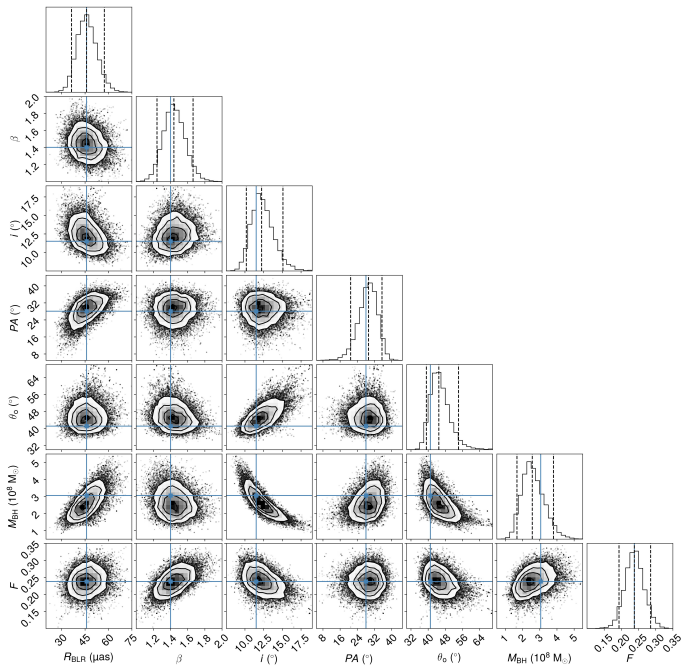

**Figure E3: Corner plot of the BLR model parameters.** 1D and 2D probability density distributions from fitting the BLR model with 7 parameters to the GRAVITY 3C 273 spectral line profile and differential phase data. The blue points show the parameters where the highest likelihood was obtained and the dashed lines are the 5, 50, and 95% quantiles.

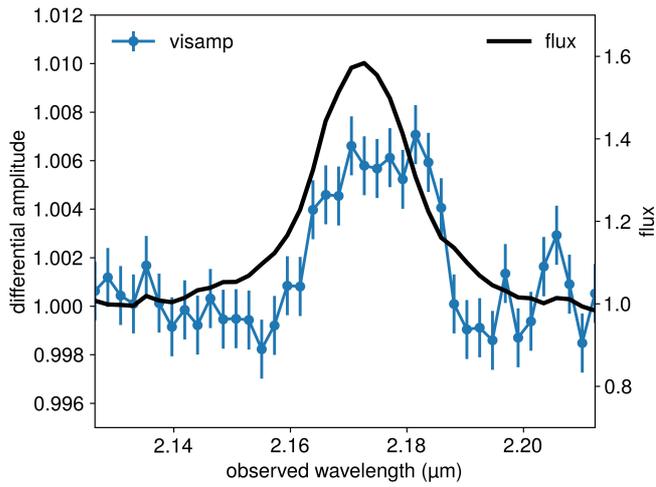

**Figure E4: Additional BLR size constraints.** 3C 273 differential visibility amplitude (blue) averaged over all epochs and between the two longest (UT4-1, UT3-1) baselines. Error bars are 1 sigma. The amplitude increases at the spectral line (black), demonstrating that the hot dust continuum (R ~ 150 μas) is much larger in size than the broad line region. This result rules out a past claim of a large region size in 3C 273[12], and is consistent with the compact size $R_{BLR}$ ~ 50 μas that we find independently from modelling interferometric phase and spectral line data.